**Nanoscale mapping of sub-gap electroluminescence from step-bunched, oxidized 4H-SiC surfaces**

*Natalia Alyabyeva, Jacques Ding, Mylène Sauty, Judith Woerle, Yann Jousseaume, Gabriel Ferro, Jeffrey C. McCallum, Jacques Peretti, Brett C. Johnson, Alistair C. H. Rowe\**

N. Alyabyeva, J. Ding, M. Sauty, J. Peretti, A. C. H. Rowe
Laboratoire de Physique de la Matière Condensée, CNRS-Ecole Polytechnique, IP Paris, 91128 Palaiseau, France

J. Woerle
Paul Scherrer Institute, Forschungsstrasse 111, 5232 Villigen, Switzerland & Advanced Power Semiconductor Laboratory, ETH Zurich, Physikstrasse 3, 8092 Zurich, Switzerland

Y. Jousseaume, G. Ferro
Laboratoire des Multimatériaux et Interfaces (UMR 5615), Université de Lyon, Université Claude Bernard Lyon 1, CNRS, 69622 Villeurbanne, France

J. C. McCallum
Physics Department, The University of Melbourne, Parkville 3052, Australia

B. C. Johnson
CQC2T, School of Engineering, RMIT University, Melbourne 3001, Australia

E-mail: alistair.rowe@polytechnique.edu

Scanning tunneling luminescence microscopy (STLM) along with scanning tunneling spectroscopy (STS) is applied to a step-bunched, oxidized 4H-SiC surface prepared on the silicon face of a commercial, n-type SiC wafer using a silicon melt process. The step-bunched surface consists of atomically smooth terraces parallel to the [0001] crystal planes, and rougher risers consisting of nanoscale steps formed by the termination of these planes. The rather striking topography of this surface is well resolved with large tip biases of the order of -8 V and set currents of magnitude less than 1 nA. Hysteresis in the STS spectra is preferentially observed on the risers suggesting that they contain a higher density of surface charge traps than the terraces where hysteresis is more frequently absent. Similarly, at 50 K intense sub-gap light emission centered around 2.4 eV is observed mainly on the risers albeit only with larger tunneling currents of magnitude equal to or greater than 10 nA. These results demonstrate that STLM holds great promise for the observation of impurities and defects responsible for sub-gap light emission with spatial resolutions approaching the length scale of the defects themselves.

## 1. Introduction

The hexagonal 4H silicon carbide (4H-SiC) polytype has become an important semiconducting material for high power electronics applications.[1] With an indirect wide bandgap (3.23 eV), 4H-SiC is host to a range of point defects responsible for sub-gap light emission from near band-edge energies out to the infrared. Some of these defects, such as the silicon vacancy[2] and di-vacancy[3] have been shown to emit single photons. Surface-related single photon emitters have also been discovered,[4,5] and these are particularly attractive for applications in quantum optics because they can be driven electrically,[5,6] stable, bright and have a high polarization visibility. However, unlike bulk emitters that emit over a small energy range, these unidentified surface states exhibit zero phonon lines (ZPL) over a large photon energy range extending from 1.55 eV to 2.45 eV. This makes their use in quantum optics applications problematic since, for example, tuning such emission into resonance with a photonic cavity to enhance spontaneous lifetimes is challenging. While optical microscopy studies of oxidized, step-bunched 4H-SiC surface indicate that the surface defect(s) responsible for single photon emission are associated with oxidation of the surface,[7,8] they are yet to be identified. Their identification is an important step towards better control and engineering of surfaces that emit at specific, desired wavelengths. With this in mind, we report here the first attempts to image sub-gap light emission on step-bunched 4H-SiC surfaces with spatial resolutions well beyond the far-field diffraction limit using a novel scanning tunneling luminescence microscope (STLM).[9]

## 2. Methods

STLM combines the selectivity of luminescence spectroscopy with the atomic-scale resolution of scanning tunnelling microscopy (STM).[10,11] A schematic diagram of the ultra-high-vacuum (UHV) experimental setup used here is shown in **Figure 1**(a). Note that the voltage $V$ is applied to the sample while the tip is held at ground as indicated. The tunneling direction of electrons can be selected according to the sign of $V$. For $V < 0$ electrons tunnel from the sample to the tip while for $V > 0$ they tunnel from the tip to the sample. In this work, n-type SiC (nitrogen doped) is used as discussed below, and the tunneling current can therefore be stabilized using the STM feedback loop[12] only for $V < 0$. Moreover, since negative voltages effectively result in hole tunneling to the sample, it is these tunnel-injected minority carriers that can recombine radiatively with majority electrons to produce the electro-luminescence which is measured when using the STM in STLM mode.

Optical access to the sample is achieved using a sample holder with a hollow center so that the electro-luminescence can be measured in transmission. The back face of the sample has a clear view to an aspherical lens (NA = 0.6, Thorlabs model 354105) integrated into the sample holder and mounted *ex situ* (i.e. prior to insertion of the sample holder into the UHV system) at a distance equal to its focal length from the emitting, front surface of the sample. This geometry is therefore suited only to samples where the substrate is transparent at the electro-luminescence wavelengths, but this in fact covers a wide range of materials and situations such as colour centres in semiconductors as described here, and also emission from near-surface hetero-structures such as quantum dots and quantum wells.[9] Since the emitted light is transmitted through the back face of the sample, its optical quality is important. In the samples studied here the back face is polished which can lead to light loss via total internal reflection in the high index substrate. An optically rough surface can be used to reduce losses due to total internal reflection, but this will also destroy the polarization state of the electro-luminescence which may be of interest. After refocusing through the lens in the sample mount, the optical beam propagates as a collimated beam through a UHV viewport and is re-focused by a second lens into a 600 μm core, multimode optical fiber. This optical fiber can then be coupled to an analysis system such as a spectrometer or a photon correlation setup. In the work described here, the output is coupled to a short focal length monochromator onto which a camera sensitive to visible wavelength is attached in order to measure electro-luminescence spectra.

In addition to its use in STLM mode, scanning tunneling spectroscopy (STS) is also reported here. While this can give access to the joint tip/sample density of states, as well as information related to possible inelastic tunneling through vibrational states,[12] it is used here only to report the root-mean-square amplitude of the lock-in current, denoted LIA, as a function of the tunneling bias. This measurement is achieved by temporarily switching off the STM feedback loop and scanning the bias over a given range. A small (15 mV) AC voltage component at 10 kHz is added to the tip bias, and the amplitude of the resulting current at this frequency is known as the LIA.

To summarize the experimental methods used here, as the tip is scanned across the surface, it can be programmed to stop at particular points for a specified period of time during which either the electrical (STS) or optical (STLM) properties can be probed and correlated with the surface topography. In this way nanoscale maps of the local electrical and optical properties can be built up.

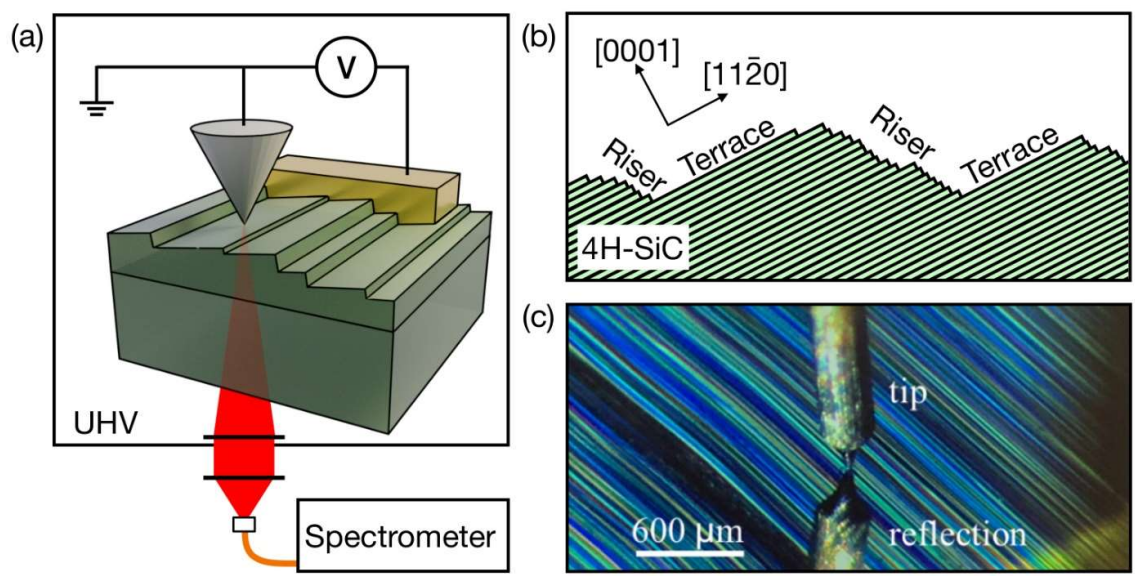

**Figure 1.** (a) A schematic diagram of the UHV STM used here. In STLM mode a voltage is applied to the sample surface via the evaporated counter electrode shown in gold while the tip is held at ground. An aspherical lens in UHV captures the transmitted electro-luminescence (red) emitted from the step-bunch 4H-SiC surface (green). After propagation through a window in the UHV system a second aspherical lens re-focusses the electro-luminescence into a multimode fiber (orange) and the light is ported to a spectrometer. (b) The step-bunched surface resulting from the silicon melt process applied to a 4° miscut, commercial 4H-SiC wafer as described elsewhere.[8,13] (c) The step-bunched surface efficiently diffracts white light yielding the colors seen in the photograph. Also visible is the STM tip and its reflection, as well as the top edge of the gold counter electrode.

With regards to the sample preparation, the silicon melt process followed by oxidation outlined in detail elsewhere[13] is applied to a commercial 4H-SiC wafer doped with nitrogen ($N_D = 4 \times 10^{15}$ cm$^{-3}$) which is miscut 4° from the (0001) basal plane. After the silicon melt process, the surface reconstructs into a step-bunched configuration indicated in the schematic diagram in Figure 1(b). The typical heights of the resulting macrosteps range from 10 nm to 100 nm, and the spatial period varies from a fraction of a micron to several microns. These macrosteps are made up of smooth terraces and rough risers containing the terminations of the [0001] crystal planes. The macrosteps are visible to the naked eye as a series of near parallel lines traversing the wafer (see Figure 1(c)). After the silicon melt process the surface was treated with HF to remove the oxide and then subsequently re-oxidized at 800 °C for 10 minutes in an $O_2$ atmosphere.[4] This process effectively oxygen terminates the surface (with a very thin $SiO_2$ layer) and gives rise to the optically-active single photon emitting defects mentioned above. The thin oxide minimizes the resistance to the STM tunneling current. Also seen in Figure 1(c) is the counter electrode in the bottom, right-hand corner, that is thermally evaporated onto the surface after the silicon melt and oxidization. Defect mapping and confocal optical microscopy indicate not only that the interface defect density is about three times higher on the risers than on the terraces, but that the highest density of single photon emitters is found preferentially on the risers rather than the terraces.[8]

## 3. Results and discussion

### 3.1. STM topography images

**Figure 2**(a) shows a 300 K, zoomed out STM image of the oxidized, step-bunched surface with $V = -8$ V and a set current of -1 nA. Terraces and risers as immediately discernible on the image, and in the line profile shown in Figure 2(b) the original 4° miscut of the commercial wafer is seen on the relatively smooth terraces, while the riser are rougher and fall at different angles. The height of the macrosteps formed by the terraces and risers falls in the 10 nm to 100 nm range as expected, and a spatial period of several microns in agreement with previous studies using atomic force microscopy.[8] Figure 2(c) shows a zoomed area on a riser, and the line profile in Figure 2(d) reveals microsteps of height ranging from 0.3 nm to 15 nm which can be compared with the $a = b = 0.3$ nm and $c = 1$ nm 4H-SiC unit cell lattice parameters.[14] It is thus

possible to obtain near-atomic resolution in the z-direction of the tip motion, although laterally atomic resolution is not achieved.

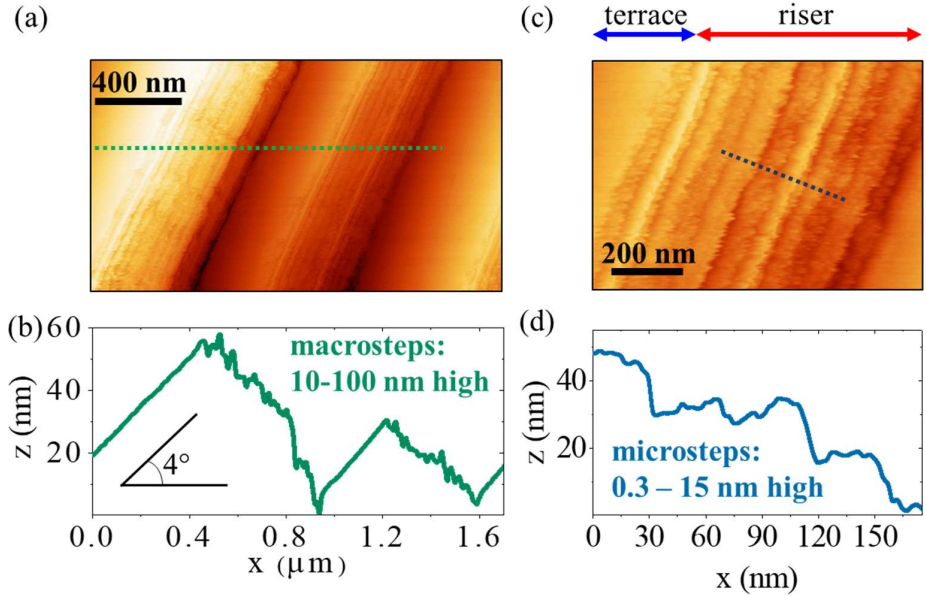

**Figure 2.** STM topography images taken at 300 K with $V = -8$ V and a current set point of -1 nA. (a) A zoomed out image of the macrosteps sketched in Figure 1(b). Terraces are large, relatively smooth areas while rises present a much higher surface roughness. The profile along the green cut line is shown in (b) where the initial 4° wafer miscut is clearly visible. (c) A zoom-in of the riser zone clearly showing the microsteps. The profile along the blue cut line is shown in (d) where steps of variable height between 0.3 nm and 15 nm are observed.

3.2 STS maps

STS is performed at 300 K where, once the STM feedback loop is switched off, $V$ is swept from -8 V to 8 V over approximately 10 s. **Figure 3** shows the two typical types of LIA versus $V$ curves obtained when doing this. At negative $V$ the LIA current rises and, depending on the position of the tip on the surface, can either exhibit a strong hysteresis between the forwards and backwards voltage sweeps as shown in Figure 3(a), or little or no hysteresis as shown in Figure 3(b). In both cases, at positive voltage no LIA current is observed as seen in the log-scale insets of Figure 3(a) and Figure 3(b)), and therefore the gap of the SiC cannot be inferred from this data. The absence of current at positive $V$ is tentatively attributed to the non-Ohmic nature of the metallic counter electrode seen in the photograph in Figure 1(c). The chrome-gold contact is evaporated directly onto the oxidized surface and may therefore exhibit rectifying or Schottky-like behavior. Indeed, for the n-type material studied here, positive $V$ corresponds to a reverse biasing of a possible Schottky contact, consistent with the absence of a measurable current.

In spite of the non-ideal nature of the counter electrode the STS measurement does reveal interesting information. STS is performed over a 32 × 32 array of points with a step size of 30 nm between points. The magnitude of the hysteresis is then estimated by calculating, at each of these 1024 points, the value of the area shaded in light blue in Figure 3(a). Since the voltage is swept at a constant rate, this area is proportional to an injected charge[15] which may be associated with the presence of charge traps at the surface. The measurement is similar to local capacitance measurements which can quantitatively reveal the density of interface states.[16,17] Here we limit ourselves to a qualitative analysis of the relative density of interface or surface states as a function of spatial position on the sample. Figure 3(c) shows the topography of the chosen zone on the sample, containing both a significant area of a riser and a terrace, each of which is identified by its surface roughness as discussed in Figure 2. Figure 3(d) shows a map whose spatial scale is identical to that of Figure 3(c). The pixel color is proportional to the stored, injected charge at each point i.e. the value of the light blue area in Figure 3(a) calculated at each point in the 32 × 32 array. Blue zones in Figure 3(d) correspond to areas where hysteresis is absent while yellow and red areas indicate significant hysteresis. The yellow to red pixels are therefore those where, according to our tentative identification of surface state charging, would indicate relatively high densities of surface defects. By comparing the topographic map shown in Figure 3(c) with this map, it is seen that the surface state density is apparently higher on riser than on the terrace, and that moreover there are significant density variations within the riser at the scale of the pixel size (30 × 30 nm$^2$). The maximum density occurs near the ridge separating the riser from the terrace. Although it is difficult to convert the electrical signal into an explicit defect density, the observation of higher densities on the risers is consistent with local deep level transient spectroscopy measurements.[8]

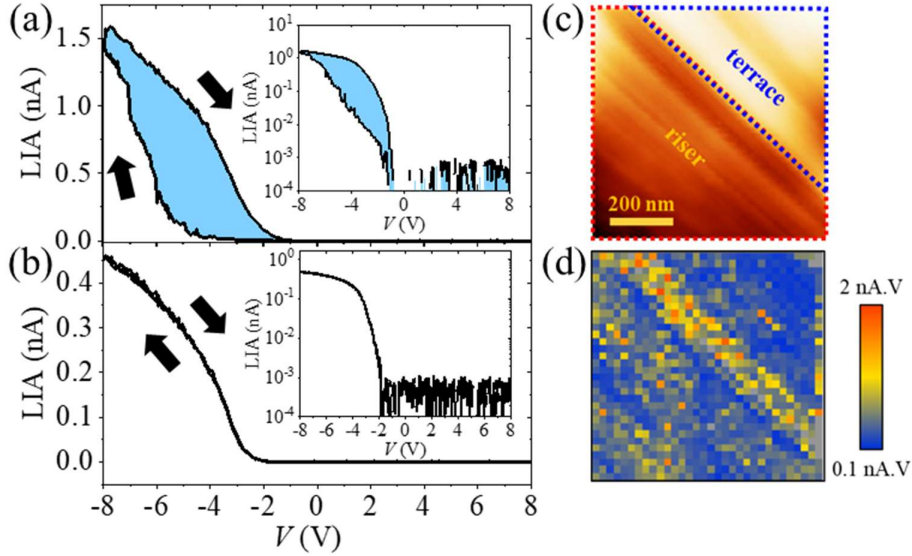

**Figure 3**. STS measurements on the oxidized, step-bunched 4H-SiC surface at 300 K. Two types of behavior are observed when sweeping forwards (-8 V to 8 V) and backwards (8 V to -8 V). Either the LIA exhibits hysteresis as shown in (a), or the hysteresis is negligibly small as in (b). The Schottky-like nature of the counter electrode on the sample stops current flow at positive $V$ where it is reversed biased. The area of the light blue shaded zone is proportional to a stored, injected charge at trap states, and varies significantly over the surface. A map of the stored, injected charge at the edge of a riser and a terrace, (c), is shown in (d). The stored charge, and hence the trap density, is systematically larger on the riser than on the terrace, but there is significant variation within the riser at the spatial resolution of the experiment (30 nm).

3.3 STLM maps

Tunnelling electro-luminescence is studied at low temperatures since a systematic increase in emitted intensity is observed as temperature is reduced. At the base temperature used here (50 K), intense electro-luminescence is observed even with only a few nA of injected tunnel current whereas several tens of nA is required at higher temperatures. This is close to the conditions where good quality topographic images can be obtained, and this facilitates studies correlating surface topography to light emission as will be discussed below. **Figure 4** shows a typical electro-luminescence spectra obtained at 50 K for $V$ = -9 V and a tunnel current of – 10 nA. The peaks in the spectrum all correspond to sub-gap emission (*c.f.* 3.23 eV for the 4H-SiC gap at 300 K). The three identified peaks shown in red, green and blue exhibit a systematic blue shift as temperature is reduced (data not shown) and peak 3 (shown in blue) exhibits the strongest increase in intensity with reductions in temperature while the other two peaks show a weaker, but similar, variation in intensity with temperature. Although the separation of the electroluminescence spectra into these components is somewhat arbitrary, this might indicate a similar origin for the emission corresponding to peaks 1 and 2, while emission corresponding to peak 3 might be due to a different defect state. However, the emission is significantly blue shifted relative to the known emission wavelengths of identified single photon emitters measured by photoluminescence. The electronic states responsible for the blue/green emission observed here are therefore not *a priori* those surface states already identified elsewhere as efficient single photon emitters.[4,5] On the other hand, a comparison of electro-luminescence maps with surface topography does reveal the that blue/green emission seen in Figure 4 arises from surface or near-surface states.

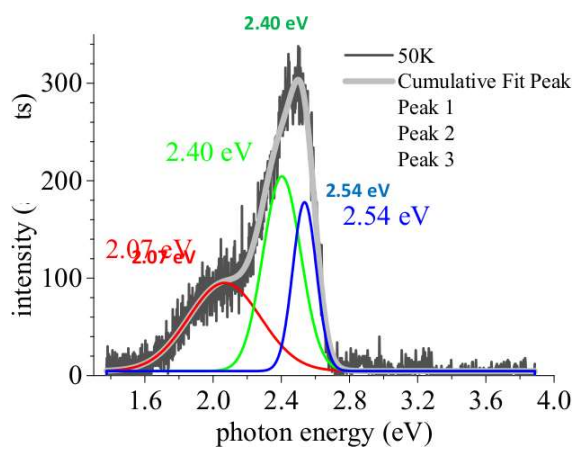

**Figure 4.** A typical tunneling electro-luminescence spectra taken at a sample temperature of 50 K, $V$ = -9 V and a tunneling current of -10 nA. The three sub-gap peaks identified in red, green and blue (and no others) are systematically observed, with varying intensity but with relatively fixed photon energies, at most points on the sample.

The large, topographic image on the left in **Figure 5** is a 1×1 μm$^2$ scan area showing terraces and risers. A square array of 32 × 32 points is defined at each of which the tip stops scanning and the electroluminescence spectrum is integrated for 10 s. The spectrum shown in Figure 4 is one of the spectra obtained in this way. A fitting procedure with the three Gaussian peaks shown in red, green and blue in Figure 4 is then applied to each of the spectra and the integrated intensity of each of the peaks is plotted as a 32 × 32 pixel map for each of the peaks as shown in the three maps to the right of Figure 5. A clear correlation between the surface topography and the light emission intensity is observed, with the strong light emission generally, but not exclusively, occurring on zones of the surface identified as belonging to risers. The preferential emission of light from risers is consistent with lower resolution confocal microscopy maps of photo-luminescence.[8] On the other hand, the superior spatial resolution of the STLM reveals significant variation within the riser on a length scale similar to the pixel size (30 nm) in the maps shown in Figure 5. One such example is the light triangular zone in the top, left of the topographic image (in Figure 5 on the left) within a riser where light emission at each of the three identified wavelengths is suppressed. The triple correlation between surface topography, the presence of hysteresis, and the intense emission of light, strongly suggests that nanometer-scale defects occurring principally within the risers, are responsible for the observed blue/green electro-luminescence observed below the 4H-SiC gap.

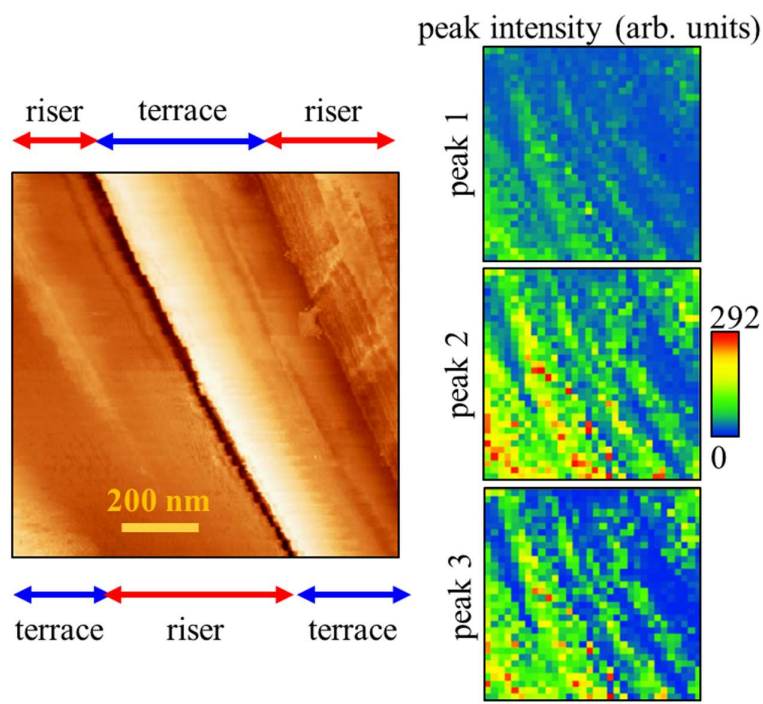

**Figure 5.** STLM of the oxidized, step-bunched 4H-Si surface at 50 K. Shown left is the topographic STM image obtained with V = -9 V and a tunneling current of -10 nA. Shown right are maps of the integrated intensities of the three peaks shown in the spectrum of Figure 4 obtained with the same current and voltage settings. A clear correlation between the topography and the light emission intensity is observable, with the light being emitted preferentially from the surface zones identified as risers.

## 4. Conclusions and future work

A combination of STM, STS and STLM has been applied to the oxidized, step-bunched surface of a n-type 4H-SiC wafer. Clear correlations between the surface topography, the hysteresis observed in the STS characteristics, and the intensity of sub-gap light emission under tunnel injection conditions, suggests that the observed blue/green electro-luminescence results from electronic transitions associated with nanoscale defect states mainly present on risers. Given that previous observations of single photon emission from the risers on this type of surface typically occur in the red to infrared part of the spectrum, it seems unlikely that the observed blue/green emission originates from the same states. While it is not yet possible to identify the details of the electronic states responsible for the emission observed here, this work demonstrates that this combination of local probe techniques, with their natural spatial resolution adapted to light emission from highly localized states, may prove to be a useful tool in their identification. This will particularly be the case when the spatial resolution of the STLM is pushed to its limit, and when the method is combined with the time resolution available in a photon correlation setup.

**Acknowledgements**

This work was partially supported by the French National Research Agency (ANR, grant TRAMP N° ANR-17-CE24-0040-01) and by the Simons Foundation (Grant N° 601944 MF).

**References**

[1] H. Lee, V. Smet, R. Tummala. *IEEE J. Emerging and Selected Topics in Power Electronics* **2019**, *8*, 239.

[2] M. Widmann, S.-Y. Lee, T. Rendler, N. T. Son, H. Fedder, S. Paik, L.-P. Yang, N. Zhao, S. Yang, I. Booker, A. Denisenko, M. Jamali, S. Ali Momenzadeh, I. Gerhardt, T. Ohshima, A. Gali, E. Janzén, J. Wrachtrup, *Nat. Materials* **2015**, *14*, 164.

[3] D. J. Christle, A. L. Falk, P. Andrich, P. V. Klimov, J. Ul Hassan, N. T. Son, E. Janzén, T. Ohshima, D. D. Awschalom, *Nat. Materials* **2015**, *14*, 160.

[4] A. Lohrmann, S. Castelletto, J. R. Klein, T. Ohshima, M. Bosi, M. Negri, D. W. M. Lau, B. C. Gibson, S. Prawer, J. C. McCallum, and B. C. Johnson, *Appl. Phys. Lett.* **2016**, *108*, 021107

[5] A. Lohrmann, N. Iwamoto, Z. Bodrog, S. Castelletto, T. Ohshima, T. J. Karle, A. Gali, S. Prawer, J. C. McCallum, B. C. Johnson, *Nat. Comms.* **2015**, *6*, 7783.

[6] M. Widmann, M. Niethammer, T. Makino, T. Rendler, S. Lasse, T. Ohshima, J. Ul Hassan, N. T. Son, S.-Y. Lee, J. Wrachtrup, *Appl. Phys. Lett.* **2018**, *112*, 231103.

[7] B. C. Johnson, J. Woerle, D. Haasmann, C. T. -K. Lew, R. A. Parker, H. Knowles, B. Pingault, M. Atature, A. Gali, S. Dimitrijev, M. Camarda, J. C. McCallum, *Phys. Rev. Appl.* **2019**, *12*, 044024.

[8] J. Woerle, B. C. Johnson, C. Bongiorno, K. Yamasue, G. Ferro, D. Dutta, T. A. Jung, H. Sigg, Y. Cho, U. Grossner, M. Camarda, *Phys. Rev. Mat.* **2019**, *3*, 084602.

[9] W. Hahn, J.-M. Lentali, P. Polovodov, N. Young, S. Nakamura, J. S. Speck, C. Weisbuch, M. Filoche, Y.-R. Wu, M. Piccardo, F. Maroun, L. Martinelli, Y. Lassailly, J. Peretti, *Phys. Rev. B* **2018**, *98*, 045305.

[0] K. Kuhnke, A. Kabakchiev, W. Stiepany, F. Zinser, R. Vogelgesang, K. Kern, *Rev. Sci. Instr.* **2010**, *81*, 113102.

[1] B. Doppagne, M. C. Chong, F. Lorchat, S. Berciaud, M. Romeo, H. Bulou, A. Boeglin, F. Scheurer, G. Schull, *Phys. Rev. Lett.* **2017**, *118*, 127401.

[2] J. Chen, *Introduction to Scanning Tunneling Microscopy Third Edition*. *Vol. 69*, Oxford University Press, USA, **2021**.

[3] V. Soulière, , D. Carole, M. Camarda, J. Wörle, U. Grossner, O. Dezellus, G. Ferro, *Mat. Sci. Forum* **2016**, *858*, 1662.

[4] F. Bechstedt, P. Kackell, A. Zywietz, K. Karch, B. Adolph, K. Tenelsen, J. Furthmuller, *Phys. Stat. Sol.(b)* **1997**, *202*, 35.

[5] Y.Fujino, K. Kita, *J. Appl. Phys.* **2016**, 120, 085710.

[6] K. Yamasue, Y. Cho, *Elec. Reliability* **2021**, *126*, 114284.

[7] K. Yamasue, Y. Cho, *Elec. Reliability* **2022**, *135*, 114588.

**Table of contents**

N. Alyabyeva, J. Ding, M. Sauty, J. Woerle, Y. Jousseaume, G. Ferro, J. C. McCallum, J. Peretti, B. C. Johnson, A. C. H. Rowe*

**Nanoscale mapping of sub-gap electroluminescence from step-bunched, oxidized 4H-SiC surfaces**

Scanning tunneling electro-luminescence microscopy reveals sub-gap light emission principally from the risers of an oxidized, step-bunched 4H-SiC surface with nanoscale spatial resolution well beyond what is possible with a confocal optical microscope.

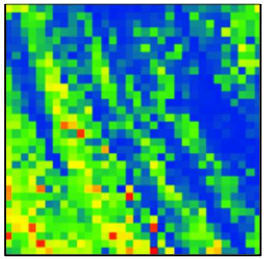